\documentclass[useAMS,usenatbib]{mn2e}
\usepackage{url,times,graphicx,amsmath,amsfonts,amssymb,aas_macros,color,epsfig,varioref,subfigure,comment}

\newcommand{\Tab}[1]{Table~\ref{#1}}
\newcommand{\Sec}[1]{Section~\ref{#1}}
\newcommand{\Eq}[1]{Eq.~(\ref{#1})}
\newcommand{\Fig}[1]{Fig.~\ref{#1}}
\newcommand{\beq}{\begin{equation}}
\newcommand{\eeq}{\end{equation}}
\newcommand{\hMpc}{{\ifmmode{h^{-1}{\rm Mpc}}\else{$h^{-1}$Mpc}\fi}}
\newcommand{\hGpc}{{\ifmmode{h^{-1}{\rm Mpc}}\else{$h^{-1}$Gpc}\fi}}
\newcommand{\hkpc}{{\ifmmode{h^{-1}{\rm kpc}}\else{$h^{-1}$kpc}\fi}}
\newcommand{\hMsun}{{\ifmmode{h^{-1}{\rm {M_{\odot}}}}\else{$h^{-1}{\rm{M_{\odot}}}$}\fi}}
\def\hMpc{$h^{-1}\,{\rm Mpc}$}
\def\hkpc{$h^{-1}\,{\rm kpc}$}
\def\LCDM{\ensuremath{\Lambda}CDM}
\def\ude{uDE}
\def\cde{cDE}

\title[Coupled and uncoupled quintessence models II: Galaxy clusters]
{Hydrodynamical simulations of coupled and uncoupled quintessence models II: Galaxy clusters}
\author[Carlesi et al.]
{Edoardo Carlesi,$^{1}$
\thanks{E-mail: edoardo.carlesi@uam.es}
Alexander Knebe,$^{1}$ Geraint F. Lewis,$^{2}$ Gustavo Yepes$^{1}$
\\
$^{1}$Departamento de F\'isica Te\'orica,
Universidad Aut\'onoma de Madrid, 28049, Cantoblanco, Madrid, Spain\\
$^{2}$Sydney Institute for Astronomy, School of Physics, 
The University of Sydney, NSW 2006, Australia}

\begin{document}

\date{Accepted XXXX . Received XXXX; in original form XXXX}

\pagerange{\pageref{firstpage}--\pageref{lastpage}} \pubyear{2013}

\maketitle

\label{firstpage}


\begin{abstract}
 We study the $z=0$ properties of clusters (and large
  groups) of galaxies within the context of interacting and
  non-interacting quintessence cosmological models, using  
  a series of adiabatic SPH simulations.
  Initially, we examine the  average properties of groups and clusters,
  quantifying their differences in \LCDM, uncoupled Dark Energy (\ude) 
  and coupled Dark Energy (\cde) cosmologies.
  In particular, we focus upon radial profiles of the gas density, 
  temperature and pressure, 
  and we also investigate how the
  standard hydrodynamic equilibrium hypothesis holds in quintessence
  cosmologies. While we are able to confirm previous
  results about the distribution of baryons, we also find that the main
  discrepancy (with differences up to $20\%$) can be seen in cluster pressure
  profiles. 
  We then switch attention to individual structures,
  mapping each halo in quintessence cosmology to its \LCDM\
  counterpart.  We are able to identify a series of small
  correlations between the coupling in the dark sector and
  halo spin, triaxiality and virialization ratio.
  When looking at spin and virialization of dark matter haloes,
  we find a weak ($5\%$) but systematic deviation 
  in fifth force scenarios from \LCDM.

\end{abstract}

\section{Introduction}

Galaxy clusters are the largest bound objects in the Universe,
with properties arising from the complex interplay between large scale gravitational dynamics and gas physics. 
For this reason, they provide a unique laboratory for probing
cosmological models on astrophysical scales, and hence to
constrain the nature of dark energy
\citep[see e.g.][]{Samushia:2008, Abdalla:2010, Carlesi:2011, 
DeBoni:2011, Baldi:2011, Allen:2011, Kogan:2012}. 
Due to the intrinsic complexity of the processes involved,
to gain theoretical insight into the formation and evolution of galaxy clusters, 
we have to employ computationally expensive hydrodynamical $N$-body simulations \citep[see][for a comprehensive review]{Kravtsov:2012}, 
and in recent years this approach has been successfully used to describe a large number of observational properties such as
X-ray temperatures, gas fractions, Sunyaev-Zel'dovich effect and pressure profiles 
\citep{Nagai:2007, Croston:2008, Arnaud:2010, Sembolini:2013}.

In an initial study \cite{Carlesi:2013a} (hereafter Paper I) 
we studied the relation between haloes and their environment,
in this work we turn to basic properties of galaxy clusters in the framework
of interacting and non-interacting quintessence cosmologies; 
such cosmologies have been developed to solve the fine-tuning problems of \LCDM~\citep[see][]{Wetterich:1995, 
Caldwell:1998, Zlatev:1999, Amendola:2000, Mangano:2003} and
their observational properties have been constrained in the background and linear regime
\citep{Amendola:2003, Pettorino:2012, Chiba:2013}, as well as in the 
highly non-linear regime by means of $N$-body simulations \citep{Maccio:2004, 
Nusser:2004, Baldi:2010, Baldi:2011, Li:2011, Baldi:2012, Carlesi:2012}.
In this paper, we will further examine our cosmological simulations, 
including 
 standard \LCDM, a free quintessence model with a Ratra-Peebles \citep{Ratra:1988}
self interaction potential (\ude, uncoupled
Dark Energy) and three quintessence models interacting with the dark matter sector 
(coupled Dark Energy, \cde033, \cde066 and \cde099).
The latter set of \cde\ models all implements a Ratra-Peebles scalar field potential 
and differ in the value of the coupling parameter $\beta_c$ only. 

Our aim is to establish links between this class of models and a set of observable
properties of galaxy clusters, firstly grouping the clusters of galaxies in each simulation
into homogeneous samples and link their properties to the cosmological framework.
We also  focus on individual structures, cross-correlating them across the different
simulations and understanding how these dark energy models influence
their properties on an object-by-object basis. 
In practice, this will reveal how structures forming from the same initial conditions, and hence
in similar environments, are affected by the global cosmological model. 

The paper is structured as follows: In \Sec{sec:simu} we will briefly introduce the physics
of the models as well as their implementation in an $N$-body code. \Sec{sec:clusters} 
discusses some of the most important features characterizing galaxy clusters in \ude\ and \cde\
scenarios, while in \Sec{sec:cross} we cross correlate them. 
In \Sec{sec:conclusions}  we  present a summary of our most important findings and 
outline the future directions of our work.


\section{Models and simulations}\label{sec:simu}
Here,  we  briefly review some of the general mathematical features  of the models studied and their numerical implementation. 
We refer the reader to Paper I and references therein for a more detailed discussion.

\subsection{Cosmological models}
Quintessence is a form of dark energy based on a cosmological 
scalar field, $\phi$, with a Lagrangian that takes the form:
\begin{equation}\label{eq:lagrangian}
L = \int d^4x \sqrt{-g} \left(-\frac{1}{2}\partial_{\mu}\partial^{\mu}\phi 
	+ V(\phi)+ m(\phi)\psi_{m}\bar{\psi}_{m} \right) 
\end{equation}
where we allow $\phi$ to interact with the matter field $\psi_m$ through the dark matter particles' 
mass term, $m(\phi)\psi\bar{\psi}$. 

The focus of this present  study are interacting and non-interacting quintessence models
with a so called Ratra-Peebles \citep[see][]{Ratra:1988} self interaction potential: 
\begin{equation}\label{eq:ratra}
V(\phi) = V_0\left(\frac{\phi}{M_p}\right)^{-\alpha}
\end{equation}
where $M_p$ is the Planck mass while $V_0$ and $\alpha$ 
are two constants whose values can be fixed by fitting the model to observational data
\citep[see][]{Wang:2012, Chiba:2013}.
While in \ude\ the mass term in \Eq{eq:lagrangian} is $m(\phi) = m_0$, 
with no interaction taking place in the dark sector;
in the latter class of models we assume that the masses of dark matter particles evolve according to:
\begin{equation}\label{eq:mass}
m(\phi) = m_0 \exp{\left(-\beta_c(\phi)\frac{\phi}{M_p}\right)} 
\end{equation}
causing an energy transfer from DM to DE due to the minus sign in front of the coupling.
In our simulations we have assumed a constant interaction term $\beta_c(\phi)=\beta_{c0}$.

\begin{table}
\caption{Values of the coupling and potential used for the \ude and \cde\ models.}
\label{tab:parameters}
\begin{center}
\begin{tabular}{cccc}
\hline
Model		& $V_0$		& $\alpha$ 	&  $\beta_c$ \\
\hline
\ude 	 	& $10^{-7}$ 	& $0.143$	& $-$ \\
\cde033	& $10^{-7}$ 	& $0.143$ 	& $0.033$\\
\cde066	& $10^{-7}$ 	& $0.143$ 	& $0.066$\\
\cde099	& $10^{-7}$ 	& $0.143$ 	& $0.099$\\
\hline
\end{tabular}
\end{center}
\end{table}

\subsection{$N$-body settings}
\begin{table}
\caption{Cosmological parameters at $z=0$ used in the \LCDM, \ude, \cde033, \cde066 and \cde099 simulations.}
\label{tab:cosmology}
\begin{center}
\begin{tabular}{cc}
\hline
Parameter & Value \\
\hline
$h$		& $0.7$   \\
$n$		& $0.951$   \\
$\Omega_{dm}$  	& $0.224$ \\
$\Omega_b$   	& $0.046$ \\
$\sigma_8$   	& $0.8$   \\
\hline
\end{tabular}
\end{center}
\end{table}

Implementing quintessence into a standard $N$-body solver requires a series of modifications that 
depend on the nature of the specific model.
Under the assumption of a very light scalar field, dark energy clustering can be neglected,
so that quintessence only acts at the background level,  
leading to a different rate of expansion with respect to the standard \LCDM\ case. 
While accounting for the changes in $H(z)$ is sufficient to properly simulate \ude\ cosmology, 
\cde\ models require a few additional modifications to be introduced, 
to take into account fifth force effects on the dark matter sector.

We implemented these features into the Tree-PM code \texttt{GADGET-2}
\citep{Springel:2005} following the algorithm of \cite{Baldi:2010a}.
To improve computational efficiency, $H(z)$, $m(z)$ and
$\phi(z)$ are being read from a series of user provided tables and not
calculated "on the fly", generating them using a customized version
of the Boltzmann solver, \texttt{CMBEASY} \citep{Doran:2005}.  Proper
initial conditions that take into account modified power spectra and
growth factors have been generated suitably modifying the
\texttt{N-GenIC} code, for $2\times 1024^3$ gas and dark matter particles in a $250$\hMpc\ box.
Gas physics has been simulated using the publicly available SPH solver of \texttt{GADGET-2},
smoothing over $N_{sph}=40$ nearest neighbours to obtain the continuous fluid quantities
and assuming a standard adiabatic relation $P\propto\rho^{\gamma}$ with $\gamma=\frac{5}{3}$, 
thus neglecting radiative effects. All of the non-standard implementations have
been carefully tested, to ensure that the new numerical techniques do not introduce systematic errors. 

\subsection{Halo catalogues}
Bound structures in our simulations have been identified using \texttt{AHF}
\footnote{\texttt{http://www.popia.ft.uam.es/AHF}}\citep{Gill:2004, Knollmann:2009},
which has been modified to take into account the influence of the different cosmologies. 
We use the equation 
\begin{equation} \label{eq:virial_mass_definition}
M_{\Delta} = \Delta \times \rho_{c}(z) \times \frac{4 \pi}{3} R_{\Delta}^{3}.
\end{equation}
to define $M_{\Delta}$ as the total mass enclosed within a radius containing
an average overdensity of $\Delta$ times the critical density of the universe
(which needs to be properly taken into account in each different cosmological model).

From the sample of haloes identified this way we restricted our
analysis to the virialized structures satisfying
\begin{equation}\label{eq:vir}
\frac{2K}{|U|} - 1 < 0.5
\end{equation}
where $K$ is the kinetic and $U$ the potential energy \citep{Prada:2012}. We therefore
ensure that unrelaxed structures (probably undergoing major mergers) do not bias our analysis.  
Even though this can be used in combination with other criteria \citep{Maccio:2007, Prada:2012}, 
we checked that their implementation would not affect our sample and thus adopted exclusively this one.
We also mention here that we do not expect the above condition to
introduce any systematic bias into our object samples drawn from the
\cde\ simulations: even though -- as we will be discussed in
\Sec{sec:virialization} -- additional couplings in the dark sector
introduce a shift into the standard virial relation, this effect is of
the order $\approx5\%$ and thus negligible with respect to the
size of the deviations removed using \Eq{eq:vir}.
\begin{figure}
\begin{center}
\includegraphics[width=8cm]{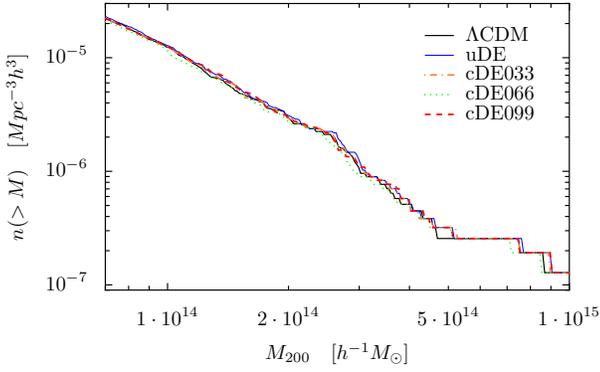}
\end{center}
\caption{\small Cluster mass function for \LCDM, \ude, \cde033, \cde066 and \cde099. 
Although the statistics in this mass regime is small, abundances are very similar for
all the models.}
\label{img:cluster_mf}
\end{figure}

\section{General properties of galaxy clusters}\label{sec:clusters}
We first study properties of structures with
mass $M>7\times10^{13}$\hMsun, which in our simulations are composed
of more than $10^5$ dark matter and gas particles. This sample
includes both clusters and large groups, and we will either use the
whole set or a smaller subset of it depending on the kind of
properties to be analyzed. In fact, due to the sharp decline of the
upper end of the cumulative halo mass function (shown in
\Fig{img:cluster_mf}), a $30\%$ reduction in the mass threshold leads
to a twofold increase in the cumulative number of objects, which can
be useful for statistical purposes.  Complementary to the cumulative
mass function (\Fig{img:cluster_mf}) we also list the total number of
clusters and large groups in each cosmology in
\Tab{tab:cluster_num}. It is evident that different models deliver
very similar results (as discussed in Paper I), although we probably need a larger computational
volume for a proper quantification of the magnitude of this effect,
minimizing the impact of cosmic variance.

\begin{table}
\caption{Number of (relaxed) galaxy clusters and large groups at $z=0$ for different mass thresholds,
found in the \LCDM, \ude, \cde033, \cde066 and \cde099
simulations.}
\label{tab:cluster_num}
\begin{center}
\begin{tabular}{ccc}
\hline
Model 	& $N(>7\times10^{13}\hMsun)$ & $N(>10^{14}\hMsun)$\\
\hline 
\LCDM 	& $338$ & $190$ \\
\ude	& $350$ & $198$ \\
\cde033 & $334$ & $193$ \\
\cde066 & $321$ & $178$ \\
\cde099 & $340$ & $193$ \\
\hline
\end{tabular}
\end{center}
\end{table}

\begin{figure*}
\begin{center}
\includegraphics[width=17cm]{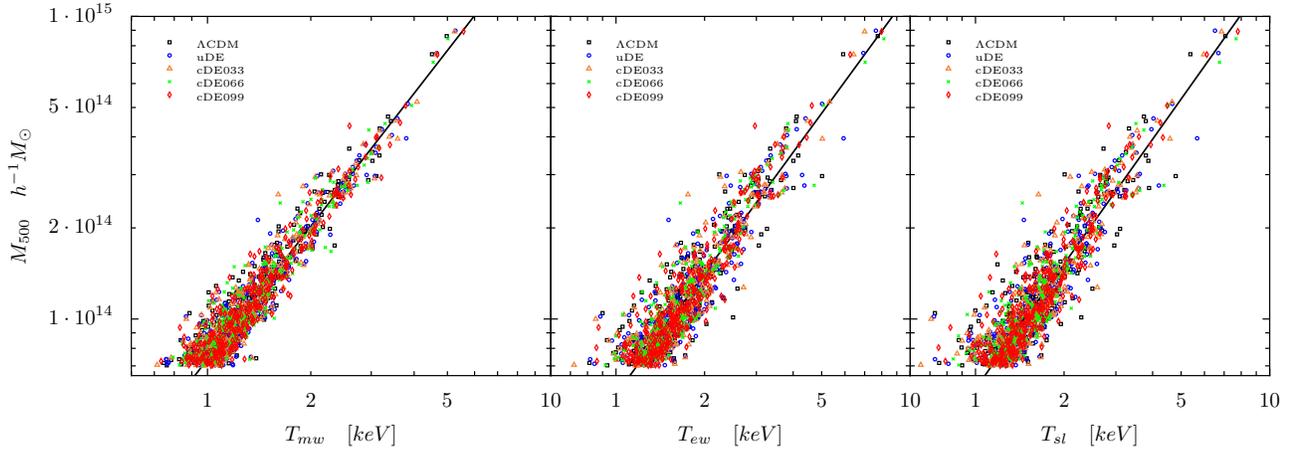}
\end{center}
\caption{\small Mass weighted, emission weighted and spectroscopic like temperatures versus 
$M_{500}$ for objects above $7\times10^{13}$\hMsun in all simulations. 
The solid black lines represents the best fit $M-T$ power law relation for \LCDM, which closely followed by all cosmological models.
}
\label{img:cluster_temp}
\end{figure*}

\begin{table*}
\caption{Best-fit values to the $M-T_{X}$, obtained fitting \Eq{eq:m_t} using $M_{500}$
versus the three temperature definitions $T_{mw}$, $T_{ew}$ and $T_{sl}$ definitions.
The $M_0$s are given in units of $10^{13}$\hMsun. 
All the models follow closely \LCDM, making this kind of relation a poor proxy for
quintessence detection.}
\label{tab:m_t}
\begin{center}
\begin{tabular}{ccccccc}
\hline
Model 	& $M_0^{mw}$ 	  & $\alpha^{mw}$   & $M_0^{ew}$    & $\alpha^{ew}$  & $M_0^{sl}$ 	& $\alpha^{sl}$\\
\hline 
\LCDM 	& $6.31 \pm 0.08$ & $1.46\pm0.03$   & $4.89\pm0.09$ &  $1.33\pm0.03$ & $5.09 \pm 0.09$  & $1.37\pm0.04$ \\
\ude 	& $6.21 \pm 0.09$ & $1.47\pm0.03$   & $4.85\pm0.07$ &  $1.33\pm0.03$ & $5.05 \pm 0.07$  & $1.38\pm0.03$ \\
\cde033 & $6.29 \pm 0.09$ & $1.46\pm0.03$   & $4.81\pm0.07$ &  $1.36\pm0.03$ & $4.95 \pm 0.08$  & $1.37\pm0.04$ \\
\cde066 & $6.31 \pm 0.08$ & $1.46\pm0.03$   & $4.96\pm0.09$ &  $1.34\pm0.03$ & $5.19 \pm 0.09$  & $1.38\pm0.04$ \\
\cde099 & $6.27 \pm 0.07$ & $1.45\pm0.03$   & $4.80\pm0.09$ &  $1.37\pm0.03$ & $5.03 \pm 0.07$  & $1.41\pm0.03$ \\
\hline
\end{tabular}
\end{center}
\end{table*}

\subsection{$T_X-M$ relation}
Cluster X-ray temperatures are an extremely important observational proxy for halo mass 
\citep{Ettori:2004, Muanwong:2006, Nagai:2007} to which they are related via a scaling relation
of the form
\begin{equation}\label{eq:m_t}
M(T_X) = M_{0} \left(\frac{T_X}{3 keV}\right)^{\alpha}
\end{equation}
where theoretical models \citep{Kaiser:1986, Navarro:1995} predict $\alpha\approx \frac{3}{2}$.
We can estimate X-ray temperatures for our simulated objects using three different definitions of $T$, namely,
the mass-weighted temperature $T_{mw}$, the emission-weighted temperature $T_{ew}$ and the spectroscopic-like temperature \citep{Mazzotta:2004}
$T_{sl}$ which reads:
\begin{equation}
T_{mw} = \frac{\sum_i m_i T_i}{\sum_i m_i}
\end{equation}

\begin{equation}
T_{ew} = \frac{\sum_i m_i \rho_i T_i \Lambda(T_i)}{\sum_i m_i \rho_i \Lambda(T_i)}
\end{equation}

\begin{equation}
T_{sl} = \frac{\sum_i m_i \rho_i T_i^{1/4}}{\sum_i m_i \rho_i T_i^{-3/4}}
\end{equation}
where $T_i$, $\rho_i$ and $m_i$ are the $i^{th}$ gas particle temperature, mass and density, 
while $\Lambda(T_i)$ is the cooling function, which we assumed to be $\propto T^{1/2}$ (thermal
Bremsstrahlung). Only particles of $T>0.5 keV$ are included in the computation of the cluster temperatures, to remove the effect of cold flows.
In \Fig{img:cluster_temp} we show the temperature mass relations for
objects larger than $7\times10^{13}$\hMsun, from which we can see that
all the models, regardless of the temperature definition, closely follow
the same $M-T$ relation of \Eq{eq:m_t}. 
This equation has been fitted using $M_{500}$ (which is closely related to the observations \cite{Sembolini:2013})
and the three different definitions of $T$ introduced before.
In the case of \LCDM\ these values
are compatible with the findings of \cite{Allen:2001, Ettori:2002, Nagai:2007} and \cite{Ventimiglia:2008}.  
It is quite clear that
the impact of quintessence on this relation is completely
negligible. Although, as we will discuss later, \cde\ models have
different effects on the properties and distribution of baryons inside
galaxy clusters, it is evident that the scaling of the X-ray
temperature with the mass is not affected in the class of quintessence
theories under investigation here.  This might be due to the
integrated nature of the relation, which conceals the details of the
underlying matter distribution of each object.
\begin{figure}
\includegraphics[width=8cm]{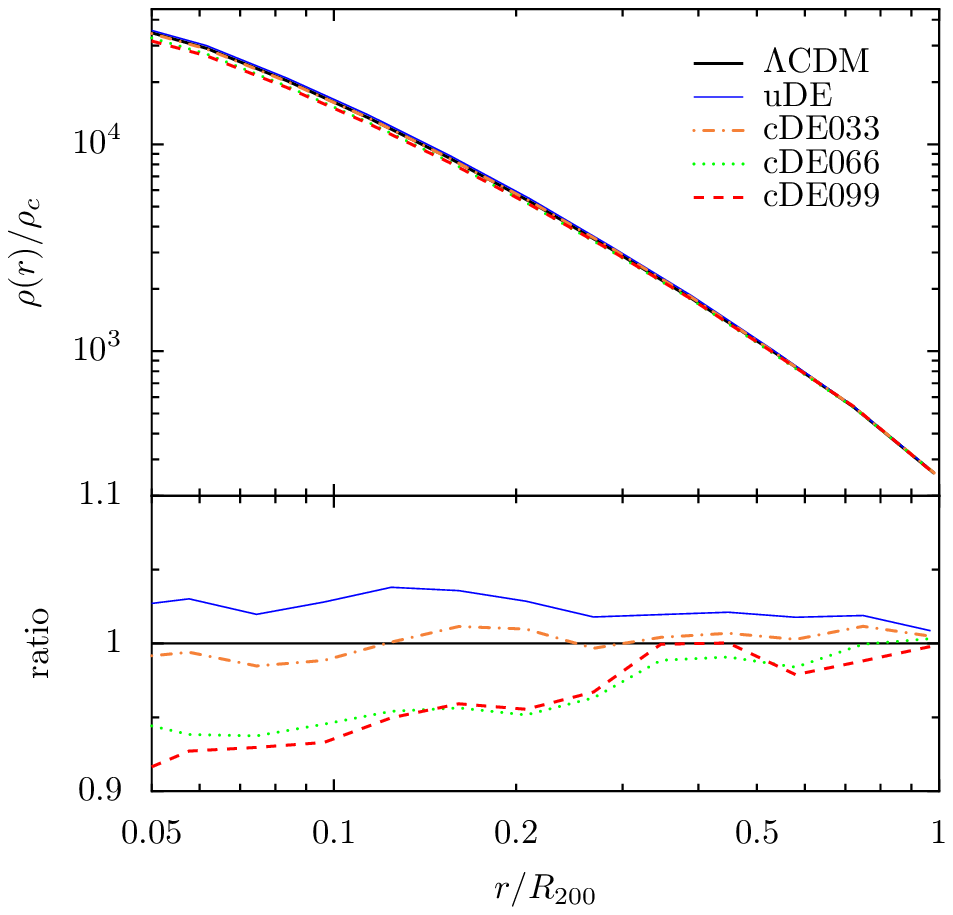}
\caption{\small Average dark matter density profile for virialized clusters above $10^{14}$ as a function of radius.
The additional interaction tends to reduce densities towards the halo center, as we can 
clearly see in \cde099 and \cde066.}
\label{img:nfw_rho}
\end{figure}

\subsection{Radial dark matter profiles}
As reported by \cite{Baldi:2010a, Li:2011}, the Navarro Frenk White (NFW) profile
\citep{NFW:1996};
\begin{equation}
\rho(r) = \frac{\rho_0}{\displaystyle \frac{r}{r_s}\left(1+\frac{r}{r_s}\right)^{2}}
\label{eq:nfw}
\end{equation}
provides a good description of the distribution of dark matter inside
virialized haloes also in the framework of \cde\ cosmologies. 
While in Paper I we already presented an analysis of density profiles for a large number of
low mass haloes, our focus here lies with the internal structure of a few, well resolved objects.  
We fit each (relaxed) halo using the radial density profiles
computed by \texttt{AHF}, which provides dark matter density for
logarithmically spaced bins assuming a spherically symmetrical
distribution.  We then compute for each halo the corresponding
goodness-of-fit $\Delta^2$ \citep{Springel:2008}, defined as
\begin{equation}\label{eq:gof}
\Delta^2 = \frac{1}{N_{bins}}\sum_{i=1}^{N_{bins}} (\log_{10}(\rho_i^{\rm (th)}) - \log_{10}(\rho_i^{\rm (num)}))^2
\end{equation}
where the $\rho_i$'s are the numerical and theoretical densities in
units of the critical density $\rho_c$ at the $i^{th}$ radial bin.

From the distribution of $\Delta^2$ (not shown here) 
we can deduce that no substantial difference can be
seen in the different cosmologies, that is, the NFW formula of
\Eq{eq:nfw} describes (on average) equally well dark matter halo
profiles in \LCDM\ as in the other (coupled) dark energy models --
something already presented in Paper I, but now extended to larger masses.

We complement this finding with \Fig{img:nfw_rho} where we show
$\rho(r)/\rho_c$ averaged over all our objects with $M>10^{14}$\hMsun\ as
a function of distance to the halo centre in units of $R_{200}$:
there, however, it is evident that the innermost regions of the
largest \cde\ clusters are associated with densities $\approx 10\%$
lower than the \LCDM\ value.  This phenomenon has also been observed
and explained -- in a different mass range -- by
\cite{Baldi:2010a}, who attributed it to the extra friction caused by the
interaction of dark energy and dark matter, which adds up to the particles' velocities causing a
small relative expansion of the halo.

\begin{figure}
\includegraphics[width=8cm]{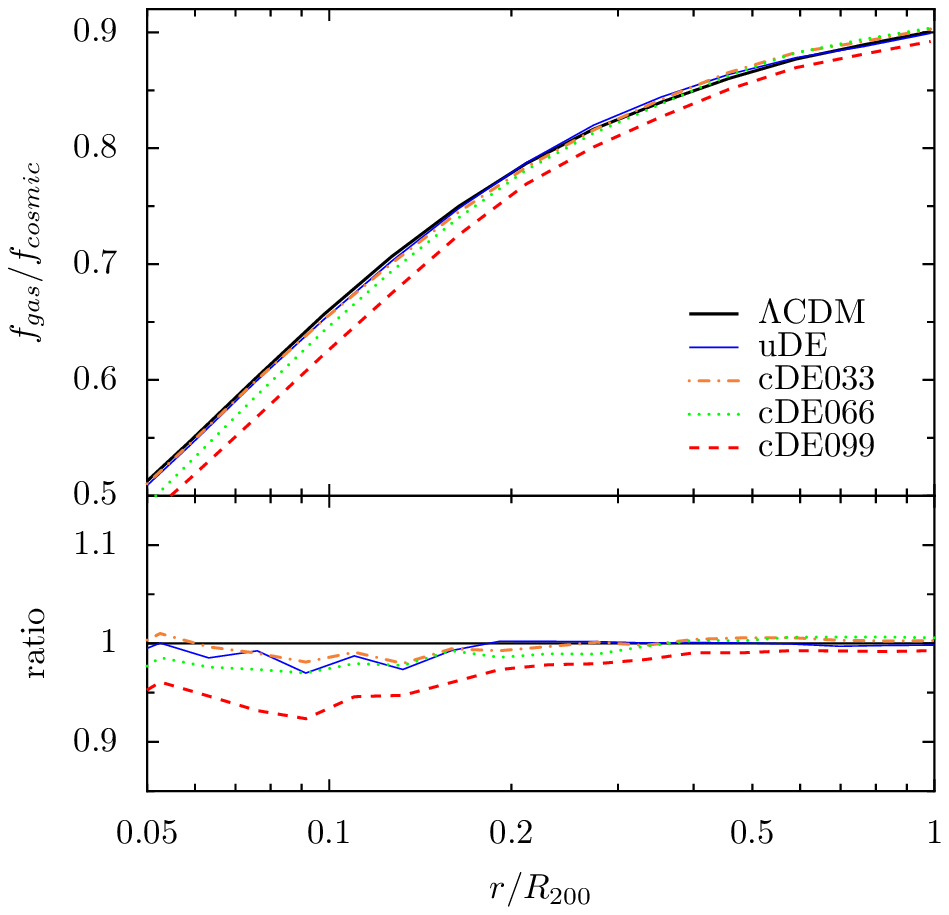}
\caption{\small Gas fraction in units of the cosmic baryon fraction
as a function of radius averaged for clusters above $10^{14}$\hMsun.
We observe that the suppression in the value of the gas fraction is stronger towards the central
regions and increases with $\beta_c$.
However, weakly interacting \cde033 
and uncoupled \ude\ are substantially indistinguishable from the standard cosmological model.}
\label{img:gas_profile}
\end{figure}

\subsection{Radial gas profiles}
Due to their large size, galaxy clusters are considered to be a "fair
sample" of the Universe, and thus should contain a fraction of baryons
close to the cosmic baryon fraction given by $\Omega_b / \Omega_m$,
where $\Omega_b$ measures the total baryonic and $\Omega_m$ the total
non-relativistic matter content.  Acting on the cosmic expansion and
thus indirectly on the collapse and formation of large structures,
we can expect quintessence to leave an imprint in the gas distribution within them.
The relation between dynamical dark energy and the
baryon content of clusters has been studied by \cite{Samushia:2008}
where they obtained constrains on the slope of the Ratra-Peebles
potential (cf. \Eq{eq:ratra}).  Gas dynamics and abundance in coupled
dark energy cosmologies have also been studied on slightly different
cosmological scales by \cite{Baldi:2010a, Baldi:2010c, Baldi:2011a}, finding a
correlation between baryon fractions and scalar field coupling to DM.

Here we add to these studies by analyzing the radial distribution of
gas and its properties like density, temperature and pressure,
focusing on structures with $M_{200}>10^{14}$\hMsun\ again, which are
composed of more than $3\times10^5$ gas and DM particles and hence
allow us to adequately resolve their internal structure.

\paragraph*{Gas fractions} In \Fig{img:gas_profile} we show
\begin{equation}
f_{gas} = \frac{M_{gas}(<r)}{M_{tot}(<r)}
\end{equation}
in units of the cosmic baryonic fraction and 
averaged over the $\approx 180$ most massive galaxy cluster in each simulation. Our
\LCDM\ results are in agreement with e.g.  \cite{Sembolini:2013}, who
found identical results for the shape of $f_{gas}(r)$ in a set of
adiabatic \LCDM\ clusters.  However we clearly observe that the net effect
of the coupling is to reduce the baryon content of the cluster in
proportion to the value of $\beta_c$. The suppression is stronger
towards the central regions of the cluster, where the average
suppression is $\approx7\%$ for \cde099 and $\approx5\%$ in \cde066,
while \cde033 and \ude\ follow closely the values of \LCDM.  At larger
radii all results tend to converge to the \LCDM\ value of $f_{gas}$,
which is slightly below the value of the cosmic baryon fraction
$\Omega_b/\Omega_m = 0.17$.  However, we must stress again that due to
the absence of radiative cooling these profiles are 
useful only as far as they allow us to provide a first
estimate of the impact of coupling in the dark sector on the (radial
distribution of the) gas content of galaxy clusters. And in that
regards, our results are in qualitative agreement with the findings of
\cite{Baldi:2010a, Baldi:2011a}, who also established a link between fifth force
and lower baryon fractions for various classes of interacting models, including
non-constant coupling models and with different types of self-interaction potentials.

This effect, called baryon segregation, was first analyzed and described in detail
in the works of \cite{Mainini:2005} and \cite{Mainini:2006}, where it was explained
in terms of the different species' infall velocity towards the centre of the potential well
in a spherical top-hat fluctuation.
In fact, this happens to be faster for DM than for baryons, since the coupling 
adds to the gravitational pull in that drives the collapse of the dark matter overdensity.
Therefore, gas particles will be relatively less abundant around the central
parts of the halo, where they are to be accreted at a slower pace, while 
their presence in the outer layers is only negligibly affected by this phenomenon.

\begin{table}
\caption{Best-fit values to \Eq{eq:betaA} for the gas density profile averaged over galaxy clusters of 
$M_{200}>10^{14}$\hMsun. The core radii $r_c$ and $r_s$ are given in units of $R_{200}$.}
\label{tab:beta}
\begin{center}
\begin{tabular}{ccccc}
\hline
Model 	& $\beta$ 	& $r_c$ 	& $r_s$ 	& $\epsilon$\\
\hline 
\LCDM 	& $0.43 \pm 0.01$ & $0.058\pm0.002$ & $0.40\pm0.05$  & $0.41\pm0.05$\\
\ude 	& $0.41 \pm 0.01$ & $0.056\pm0.002$ & $0.34\pm0.07$ &  $0.38\pm0.05$\\
\cde033 & $0.41 \pm 0.02$ & $0.053\pm0.002$ & $0.33\pm0.07$ &  $0.36\pm0.08$\\
\cde066 & $0.39 \pm 0.01$ & $0.053\pm0.003$ & $0.33\pm0.05$ &  $0.36\pm0.05$\\
\cde099 & $0.42 \pm 0.02$ & $0.064\pm0.004$ & $0.36\pm0.05$ &  $0.35\pm0.04$\\
\hline
\end{tabular}
\end{center}
\end{table}

\begin{figure}
\includegraphics[width=8cm]{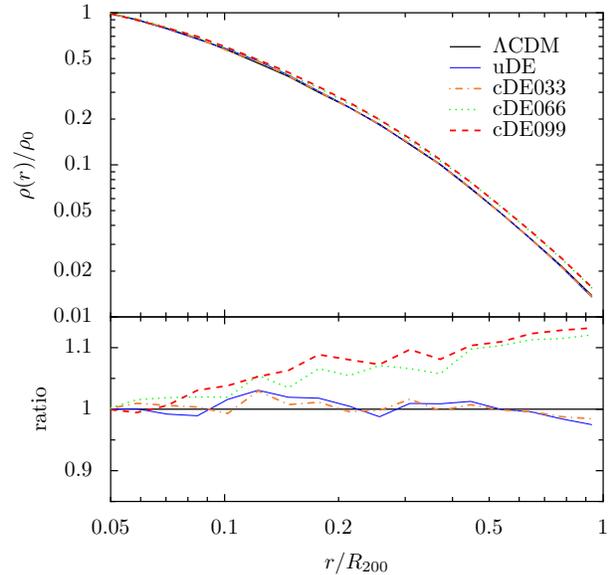}
\caption{\small Radial distribution of gas density averaged for clusters above $10^{14}$\hMsun, normalized to the
central density $\rho_0$.}
\label{img:gas_beta}
\end{figure}

\paragraph*{Density profile} After studying how the baryon fraction (which is a combination of gas
and dark matter properties) is affected we consider whether the
coupling also induces sizeable effects in the gas density profile
alone.  Under the assumption of hydrostatic equilibrium (which
holds to the same degree in both quintessence models and \LCDM\ --
as we will see in \Sec{sec:hse} below) we can derive a simple
functional form for the gas density profile \citep{Cavaliere:1976}, the
so called $\beta$\footnote{This $\beta$ must not be confused with $\beta_c$, the coupling parameter} 
model:
\begin{equation}\label{eq:beta}
\rho(r) = \rho_0 \left(1 +\left(\frac{r}{r_c}\right)^2\right)^{-\frac{3}{2}\beta}
\end{equation}
where $r_c$ is the core radius and $\rho_0$ is the inner cluster
density, which is defined as $\rho(r=0.05\times R_{200})$.
Observations \citep{Vikhlinin:1999} and simulations \citep{Rasia:2004}
have shown that \Eq{eq:beta} does not simultaneously reproduce 
both the inner and outer parts of density distribution of galaxy clusters,
even though this model can still be used as a valuable theoretical
tool that captures the main characteristics of the intra-cluster
medium (ICM) \citep{Arnaud:2009}.  Hence, for a quantitative
comparison of the results for radial distribution of gas densities in
the different cosmologies we refer here to a model proposed by
\cite{Mroczkowski:2009}. This was developed for the observational
description of X-ray cluster profiles, and is based on the formula
proposed in \citep{Vikhlinin:2006}, which in turn is an extension of
the simple $\beta$ model.  Here we re-write \Eq{eq:beta} as:
\begin{equation}\label{eq:betaA}
\frac{\rho(r)}{\rho_0} = \frac{1}{\left( 1 +\frac{r^2}{r_c^2}\right)^{\frac{3}{2}\beta}} \times
\frac{1}{\left(1 + \frac{r^3}{r_s^3}\right)^{\epsilon}}
\end{equation}
where the additional multiplicative term on the right contains a new
scale radius $r_s$ and power law $\epsilon$, which capture the
departure from the standard $\beta$ model at larger radii.  We then
compute the average $\rho(r)/\rho_0$ per radial bin (in
units of $R_{200}$), again using all clusters of $M_{200}>10^{14}$\hMsun. 
We check that \Eq{eq:betaA}
holds for all the models verifying that the corresponding goodness of
fits take comparable values (analogously defined to \Eq{eq:gof}); and
in \Tab{tab:beta} we show the best-fit parameters; note that we defer
from showing the best-fit curves in \Fig{img:gas_beta} again to not
overload the plot. The parameters  do not seem to show any trend
for \cde\ and \ude\ models, except for a slightly shallower outer
slope $\epsilon$ in coupled cosmologies which can be also seen in
\Fig{img:gas_beta} where we present the averaged radial gas
distribution. We also notice that for our objects $\beta$ is
substantially lower than usually assumed ($\approx0.66$), however,
this can be easily explained by the fact that our model has two
different slopes to account for the two different regimes: this means
that, being biased by the core regions of the cluster, where the
decrease in density is shallower, $\beta$ will consequently be
smaller.

\Fig{img:gas_beta} further shows clearly that -- away from the
center of the clusters -- the \cde066 and \cde099 gas densities
increasingly diverge from the other models, up to more than $10\%$ at
the outer edge.  As discussed earlier, using the theoretical 
framework of \cite{Mainini:2005, Mainini:2006}, these models are
characterized by lower baryon fractions in the central regions
of the clusters (i.e. a smaller $\rho_0$, according to our definition)
which on the other hand converge to \LCDM, \cde033 and \ude\ in the
outer regions.  Hence, divergences in $\rho(r)$ for $r\rightarrow
R_{200}$ are explained by the small denominator $\rho_0$, enhancing
even more the gap between the predictions of coupled quintessence
cosmologies and the standard model.
\begin{figure}
\includegraphics[width=8cm]{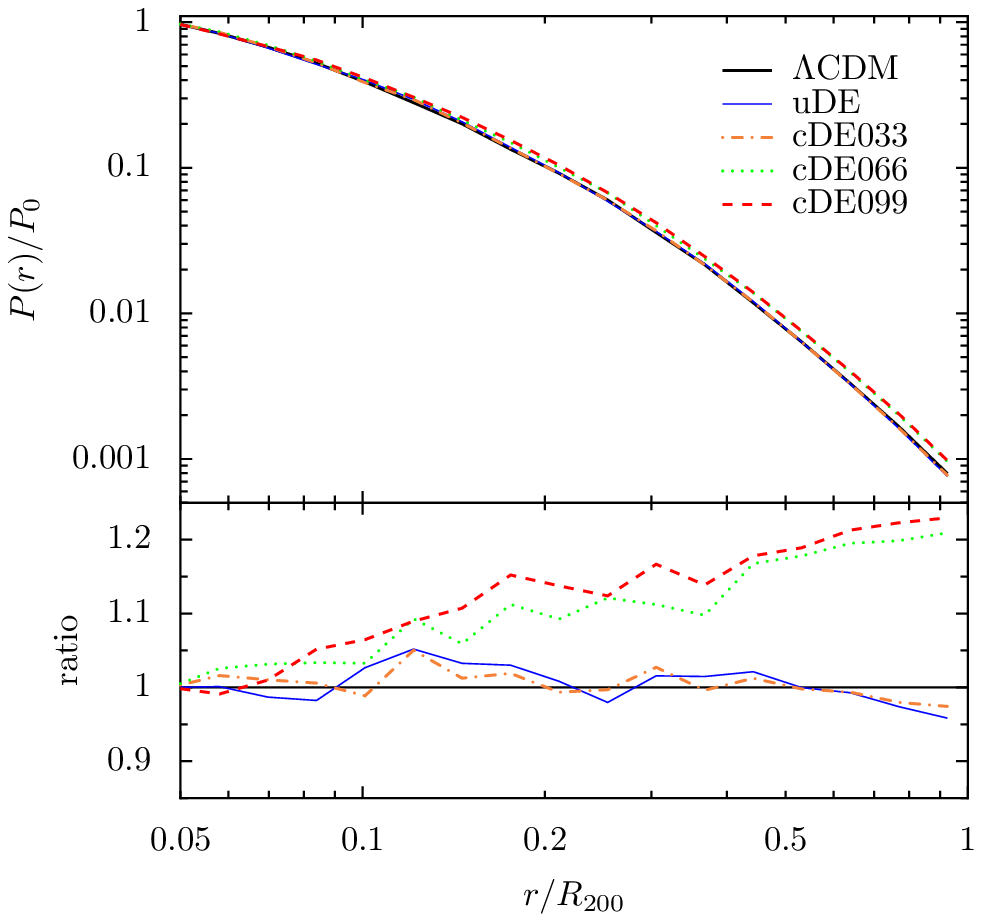}
\caption{\small Pressure profiles averaged over clusters above $10^{14}$\hMsun.}
\label{img:pressure_profile}
\end{figure}

\paragraph*{Pressure profiles} Having analyzed the properties of baryon density distributions, we now
consider the pressure profiles, which can be modeled assuming a
simple adiabatic relation of the type
\begin{equation}\label{eq:pressure}
P(r) = P_0 \rho_{gas}^{\gamma}(r)
\end{equation}
where $P_0$ is an arbitrary normalization constant (which we take to
be equal to $P(0.05\times R_{200})$), and $\gamma=5/3$ for the case of
an adiabatic gas -- as assumed in our simulations.  Using the
densities computed in the previous sub-section, it is straight-forward to obtain
the pressure profiles by using \Eq{eq:pressure}; the results are
plotted in \Fig{img:pressure_profile}. It is remarkable that the
differences among the models are now much larger (note the enlarged
range in the ratio plot), since the discrepancies observed previously
are now basically amplified by the adiabatic index $\gamma$. Again,
this effect increases towards the outer halo edge, where the ratio of
$\rho_{gas}(r)$ to the inner density $\rho_0$ is larger in \cde\
models due to the under-abundance of gas in the central regions.

Qualitatively, the shapes in \Fig{img:pressure_profile} reproduce well
the so-called universal pressure profile of galaxy clusters, which can
be reconstructed using Sunayev-Zel'dovich effect and X-ray data
\citep{Arnaud:2010, Bonamente:2012, Planck:2013}.  However, the errors
on the observational results are still larger than the spread among
the different models considered here so that for the moment it is not
possible to use these dataset to directly constrain quintessence.
Moreover, a direct comparison to the data would probably require to
relax the unrealistic assumption of completely adiabatic gas and
introduce additional effects (such as radiative cooling, star formation, 
and AGN feedback) to properly simulate the gas
physics.  In any case, it is clear that gas pressure in cluster does
represent an important probe for \cde\ cosmologies, as the non-linear
relation between gas and pressure significantly magnifies the
prediction of scarcer gas in the core regions characteristic of these
cosmological models.

\begin{table}
\caption{Best fit values to a linear relation for the outermost values of the temperature profile.
The vertical dashed line denotes the innermost excluded region, where the linear relation 
does not hold.}
\label{tab:temp}
\begin{center}
\begin{tabular}{ccc}
\hline
Model 	& $A$ 			& $B$ \\
\hline 
\LCDM 	& $1.05 \pm 0.01$ & $0.61\pm0.02$\\
\ude 	& $1.10 \pm 0.01$ & $0.61\pm0.02$\\
\cde033 & $1.10 \pm 0.01$ & $0.61\pm0.02$\\
\cde066 & $1.11 \pm 0.02$ & $0.59\pm0.01$\\
\cde099 & $1.11 \pm 0.02$ & $0.59\pm0.01$\\
\hline
\end{tabular}
\end{center}
\end{table}

\begin{figure}
\includegraphics[width=8cm]{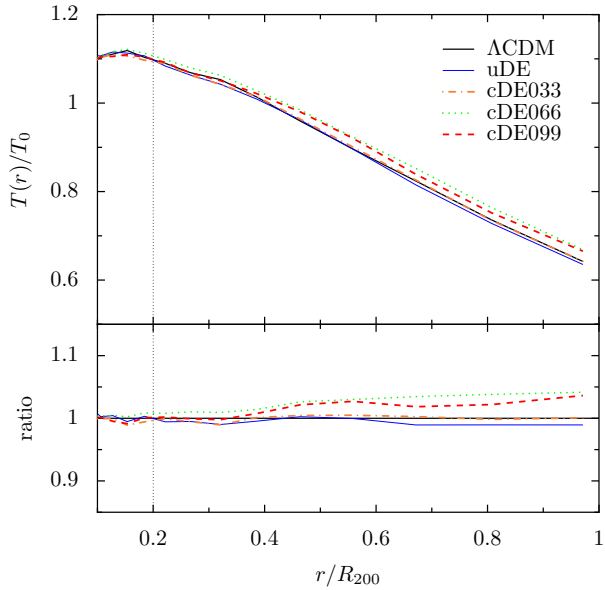}
\caption{\small Temperature profiles averaged over clusters above $10^{14}$\hMsun.}
\label{img:temp_profile}
\end{figure}

\paragraph*{Temperature profiles} Observations have shown \citep[e.g.]{Markevitch:1998, Vikhlinin:2005}
that galaxy clusters have a declining temperature towards larger
radii, in contrast with the simplest isothermal models.  The
same pattern is seen in our simulations, as the curves in
\Fig{img:temp_profile} show, and is in qualitative agreement with the
findings of \cite{Vikhlinin:2006, Arnaud:2010, ABaldi:2012}.
However, it is known that adiabatic SPH simulations fail to reproduce
the inner cool core of galaxy clusters \citep{Kravtsov:2012} up to a
value of $\approx0.2\times R_{200}$; this point is marked by a vertical dotted line in
\Fig{img:temp_profile}.

Following \cite{DeGrandi:2002} and \cite{Leccardi:2008} we model 
the outer parts of galaxy clusters using a linear function
\begin{equation}\label{eq:temp}
\frac{T(r)}{T_0} = A - B\left(\frac{r}{R_{200}} - 0.2\right)
\end{equation}
where $A$ and $B$ are two free parameters and $T_0$ is the average
temperature for each cluster.  We proceed identifying all structures
above $10^{14}$\hMsun\ and fitting \Eq{eq:temp} using the gas
densities and temperatures for regions of $r > 0.2 \times R_{200}$.
The best-fit values are presented in \Tab{tab:temp} while only the
numerical results are plotted in \Fig{img:temp_profile}.  The five
profiles are very similar and the largest differences can be seen in
the strongest coupled cases of \cde066 and \cde099, where the scaled
temperature at the halo edge is $\approx5\%$ larger than in the other
models. However, all the points as well as the best-fit values are
still consistent within the error so that this small difference
is considered to be not significant.  The effect of the coupling is
thus marginal in this case, and it seems unlikely that radial
temperature measurements alone can provide a mean to distinguish
amongst at least the models considered here.

\subsection{Gas alignment to the dark matter halo}
\begin{figure}
\includegraphics[width=8cm]{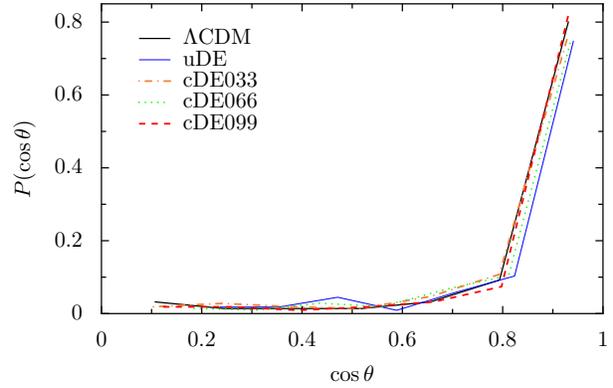}
\caption{\small Average cosine of the alignment angle between the gas and dark matter major axes, 
averaged for objects with $M_{200}>10^{14}$\hMsun. }
\label{img:gas_distribution}
\end{figure}

We now extend our study to the full 3D distribution of the gas inside
the halo, i.e. we are considering the shape of the gas particles spatial distribution.  To
this extent, we utilize the inertia tensor 
\begin{equation}\label{eq:itens_b}
  I_{ij}^{\rm gas} = \sum_{n_{\rm gas}} x_{(n),i}^{\rm gas} \ \ x_{(n),j}^{\rm gas} 
\end{equation}
where $x_{(n),i}^{\rm gas}$ is the position vector relative to the center of the
baryon mass distribution of the $n^{th}$ particle.  In the same way we
write the halo's inertia tensor
\begin{equation}\label{eq:itens_dm}
I_{ij}^{\rm dm} = \sum_{n_{\rm dm}} x_{(n),i}^{\rm dm} \ \ x_{(n),j}^{\rm dm} 
\end{equation}
which is now given by summing over dark matter particles only. We then
diagonalize the two tensors using the two largest eigenvectors
$\mathbf{e}_{1}^{h}$ and $\mathbf{e}_{1}^{b}$ -- which are the major
axes of the dark matter and baryon distribution, respectively -- in
what follows.  To check whether quintessence has an influence on
the relative spatial distribution of gas and dark matter particles we compute
\begin{equation}
\cos\theta = \frac{\mathbf{e}_{1}^{h} \cdot \mathbf{e}_1^b} {|e_1^{h} e_1^{b}|}
\end{equation}
for all clusters above $10^{14}$\hMsun\ again. The probability
distribution of $\cos\theta$ is shown in \Fig{img:gas_distribution},
where we can see that all cosmological models follow the same pattern
of close alignment between gas and dark matter distributions, although
with some scatter among the models at small angles, where $\cos\theta
\rightarrow 1$.  
We note here that our results refer to the gas properties only, and cannot
be directly compared to \cite{Lee:2010} and \cite{Baldi:2011b}, who looked
at galaxy alignment.

\begin{figure}
\includegraphics[angle=270, width=8cm]{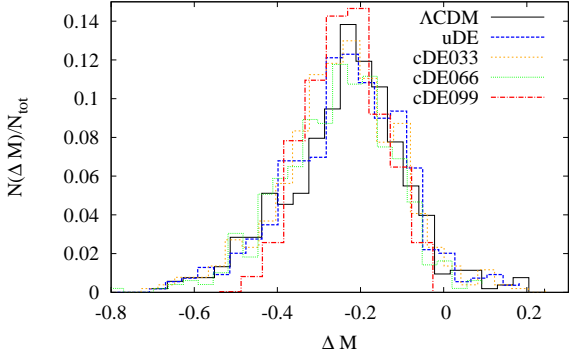}
\caption{\small Distribution of $\Delta M = (M_{HSE,200} - M_{200})/M_{200}$, the 
difference between the cluster mass estimated using the hydrostatic equilibrium assumption and the true
mass obtained in the simulations. 
We computed $M_{HSE}$ for relaxed haloes of $M>9\times10^{13}$\hMsun. 
The distribution of $\Delta M$ is peaked around $-0.12$ for all models, showing no large deviation
from the \LCDM\ pattern neither in \ude\ nor in \cde\ cosmologies.}
\label{img:hydro_mass}
\end{figure}

\begin{table}
\caption{Best fit values to a Gaussian distribution for the $\Delta M$ computed under the hypothesis of
HSE. While all the models tend to produce similar results, 
we see that \cde099 has a narrower dispersion around the peak; yet the absence of a comparable effect
in the other \cde\ models indicates that the correlation to the coupling is at best very weak.}
\label{tab:hydro_mass}
\begin{center}
\begin{tabular}{ccc}
\hline
Model 	& $\Delta M_0$ 		& $\sigma$ \\
\hline 
\LCDM 	& $0.23 \pm 0.02$ & $0.35\pm0.03$\\
\ude 	& $0.24 \pm 0.02$ & $0.37\pm0.04$\\
\cde033 & $0.22 \pm 0.02$ & $0.36\pm0.04$\\
\cde066 & $0.23 \pm 0.02$ & $0.35\pm0.03$\\
\cde099 & $0.23 \pm 0.02$ & $0.28\pm0.03$\\
\hline
\end{tabular}
\end{center}
\end{table}

\subsection{Hydrostatic equilibrium}
\label{sec:hse}
Observations of galaxy clusters usually assume hydrostatic equilibrium
(HSE) to derive their masses.  Under this hypothesis, gas and galaxies
are both in equilibrium with the binding cluster gravitational
potential \citep{Sarazin:1986, Evrard:1990, Bahcall:1994}.  However,
this assumption is not always valid and is a major source of
uncertainty when deriving observational scaling relations. Many
authors \citep[e.g.][]{Ameglio:2009, Lau:2009, Sembolini:2013} found a
systematic underestimation of cluster masses within the range
$10-25\%$ for \LCDM. This was explained by \cite{Lau:2009} and
identified as an effect driven by random gas motion that contribute to
the pressure support, which in HSE is used to estimate the mass using
the relation
\begin{equation}\label{eq:hse}
M_{HSE} (<r) = - \frac{k T_{mw} r}{G m_{H} \mu} \left(\frac{d \ln \rho}{d \ln r} + \frac{d \ln T}{d \ln r} \right).
\end{equation}
where $k$ is the Boltzmann constant, $T_{mw}$ is the mass weighted temperature, $m_{H}$ is the
hydrogen mass, $\mu$ is the hydrogen mass fraction and $\rho$ is the gas density.

We are interested in examining the impact of alternative
cosmological scenarios on the above estimation as the
effective dark matter gravitational potential is affected by the
presence of an additional interaction mediated by the dark energy.  To
accomplish this we identify relaxed clusters (as defined by
\Eq{eq:vir}) of $M_{200}>\times10^{14}$\hMsun, and compute for each one of
them the function $M_{HSE}(<r)$ using the temperature and pressure
profiles.  We can then straight-forwardly obtain an estimated total mass
$M_{HSE,200}$ simply by using its value at the halo edge $R_{200}$,
defined by \Eq{eq:virial_mass_definition}.

The distribution of the fractional difference
\begin{equation}
\Delta M = \frac{M_{HSE,200} - M_{200}}{M_{200}}
\end{equation}
with respects to the true mass as returned by the halo finder is shown
in \Fig{img:hydro_mass}, where we clearly see that this mass estimator
has an average negative bias peaked around $\Delta M_0=-0.22$ and a dispersion
$\sigma=0.34$ for all models (as shown in \Tab{tab:hydro_mass}), 
except for \cde099 which shows a slightly more pronounced
peak and a narrower dispersion around it.
However, because of the absence of such a trend in the other \cde\ models and the non-negligible 
error bars, it appears more likely that this effect is due to a statistical
fluctuation.

It is thus safe to state that \ude\ and \cde\ cosmologies are not 
responsible for any additional bias, even though the use of a larger halo sample containing more
clusters with $M_{200}>10^{15}$\hMsun\ might be needed to test whether
this statement really holds at even higher mass scales.

\section{Cross-correlated properties of galaxy clusters}\label{sec:cross}
Due to our approach of using the same random phases for all models
when generating the initial conditions for the simulations we are in
the situation of cross-identifying the same objects in all the models.
Therefore, focusing on structures forming in the same environments
whose evolution is driven by different laws, we can shed more light
into the effects of cosmic evolution on properties of individual
objects and describe how they change when switching from one model to
the other. Or put differently, while in the previous sections we
primarily engaged in studying distribution functions, we are now
directly testing the influence of our models onto individual objects.

The cross-correlation was performed matching 
every \LCDM\ halo with $M_{200}>7 \times 10^{13}$ with its counterpart,
i.e. 338 haloes were sought in the other models
(cf. \Tab{tab:cluster_num}). But this mass cut was only applied to the
\LCDM\ haloes and we were hence able to cross-match every of those
\LCDM\ haloes. To actually cross-identify objects we used a halo
matching technique that correlates those \LCDM\ haloes to the halo
catalogue of the other models by examining the particle ID lists and
maximizing the merit function $C=N_{\rm shared}^2/(N_{1}N_{2})$, where
$N_{\rm shared}$ is the number of particles shared by two objects, and
$N_{1}$ and $N_{2}$ are the number of particles in each object,
respectively \citep[e.g.][]{Knebe:2013}.

\begin{table*}
\begin{center}
\caption{
Average of the model to \LCDM\ ratio for a series of cross correlated objects with their
dispersion. $M_{200}$ is the cross correlated halo mass, $|U|/(2K)$ the ratio of the virialization of each object,
$T_{mw}$ the mass weighted temperature, $f_{gas}$ the gas content as a fraction of the total mass, $\lambda$
the spin parameter and $t$ the triaxiality parameter.
Even though the scatter is significant, 
we can see a correlation of $\lambda$, $t$ and virialization to the dark energy coupling, while the
other parameters' average are largely independent of the model.}
\label{tab:averages}
\begin{tabular}{ccccc}
\hline
Parameter	& \ude 		& \cde033 	& \cde066 	& \cde099 \\
\hline 
$M_{200}$	&$1.03\pm0.01$	&$1.03\pm0.03$	&$0.99\pm0.02$	&$1.04\pm0.02$\\
$|U|/(2K)$	&$0.997\pm0.003$&$1.005\pm0.005$&$1.015\pm0.005$&$1.05\pm0.01$\\
$T_{mw}$	&$0.998\pm0.002$&$1.015\pm0.004$&$0.991\pm0.006$&$1.03\pm0.01$\\
$f_{gas}$	&$1.029\pm0.003$&$1.001\pm0.001$&$1.002\pm0.002$&$0.991\pm0.006$\\
$\lambda$	&$1.02\pm0.04$	&$1.06\pm0.03$	&$1.05\pm0.03$	&$1.10\pm0.04$\\
$t$		&$1.01\pm0.04$	&$1.03\pm0.02$	&$1.06\pm0.03$	&$1.05\pm0.02$	\\
\hline
\end{tabular}
\end{center}
\end{table*}

\begin{figure*}
\begin{center}
\includegraphics[width=16cm]{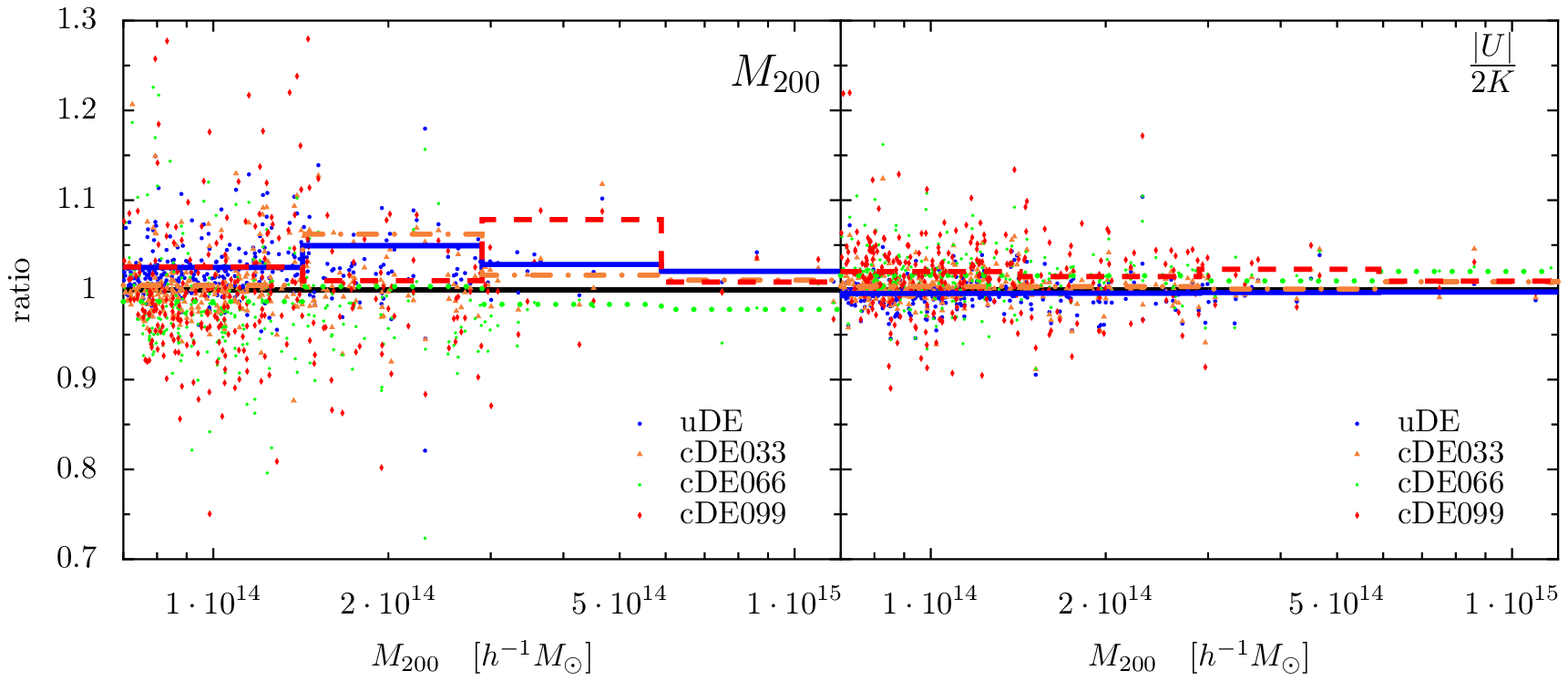} \\
\includegraphics[width=16cm]{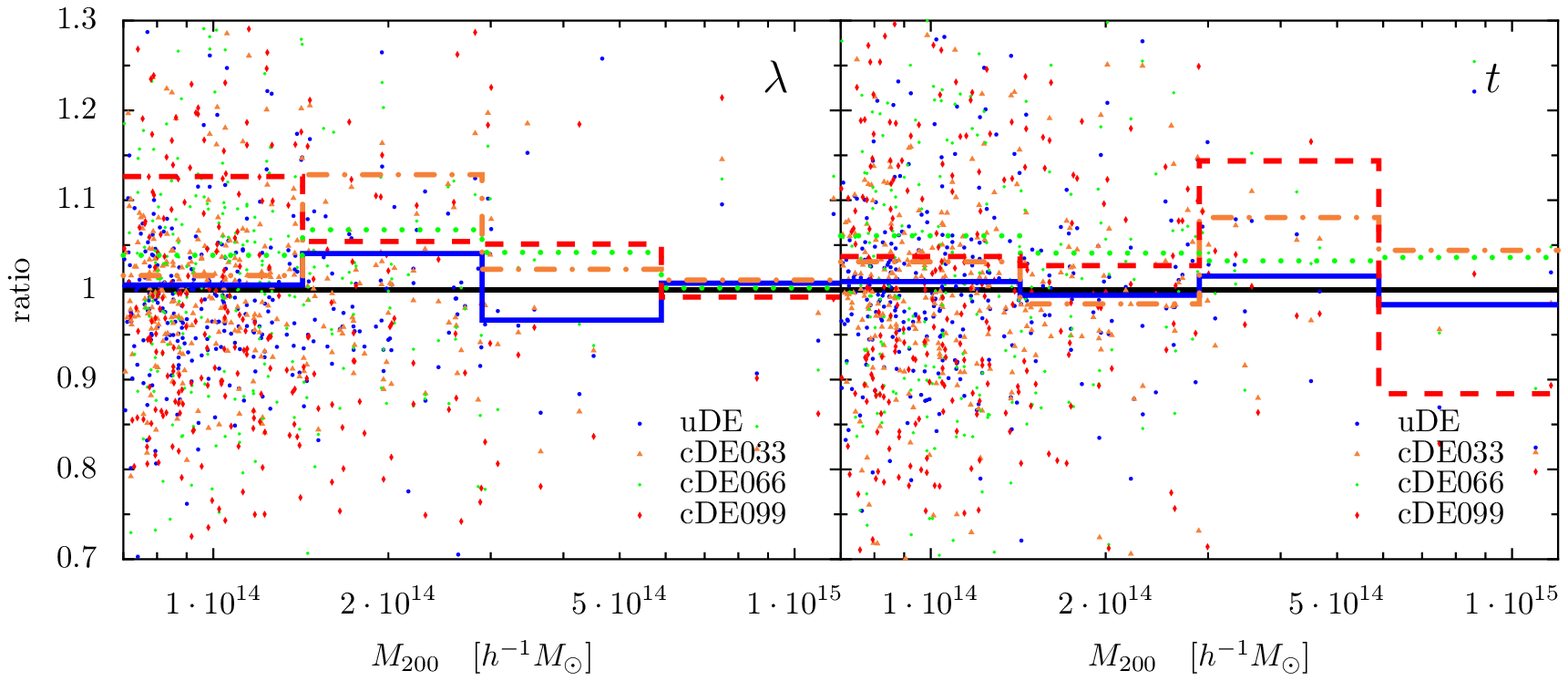} \\
\includegraphics[width=16cm]{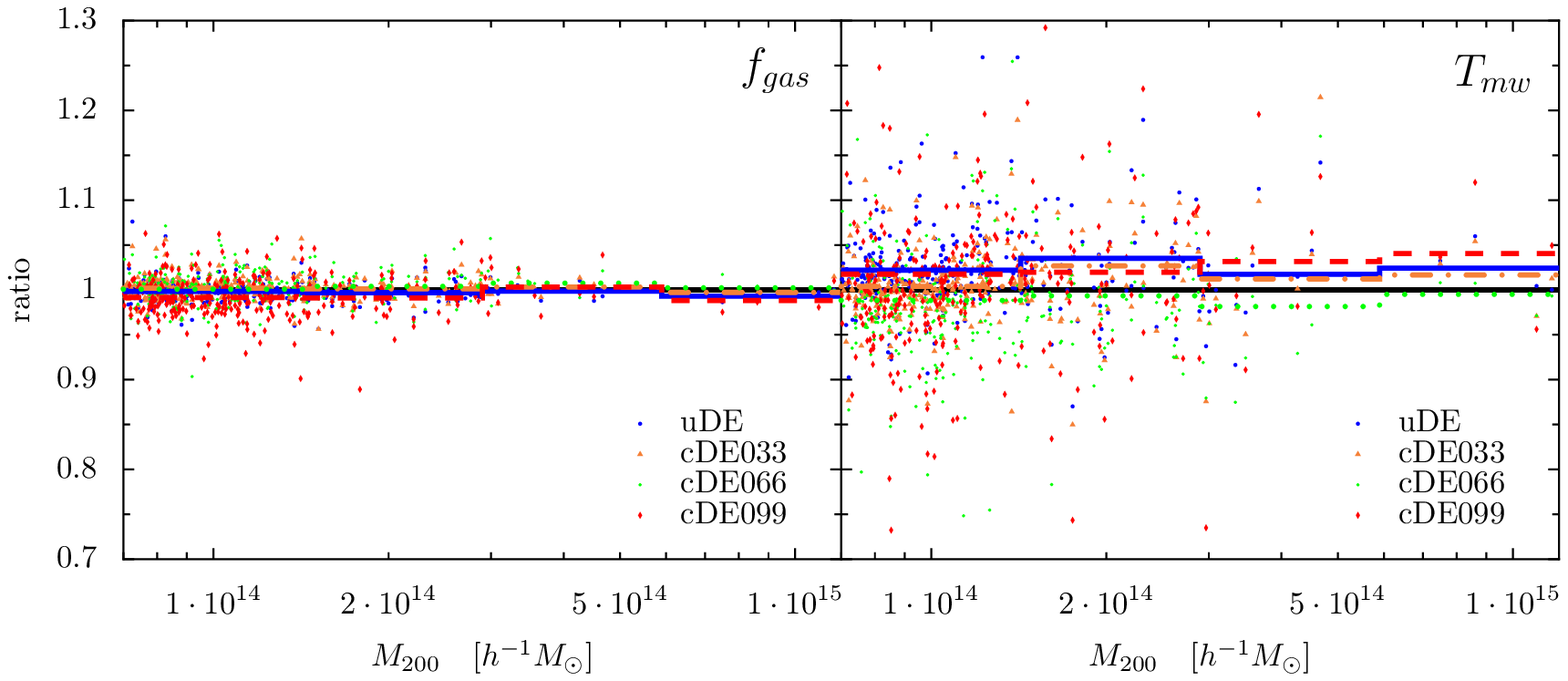} \\
\caption{\small Halo and gas properties in quintessence models. Each dot represents the value of 
the ratio of the parameter in \cde\ or \ude\ to its cross correlated structures in \LCDM.
Upper panels: halo mass (left) and virialization (right). Central panel: halo spin (left) and 
triaxiality (right). Lower panel: gas fraction (left) and mass-weighted temperature (right).
}
\label{img:cross_correlation}
\end{center}

\end{figure*}

For each pair we then compare $M_{200}$, virialization, spin
parameter, mass weighted temperature and gas fraction. The results are
all summarized in \Fig{img:cross_correlation} and \Tab{tab:averages}.
Although most of these distribution look quite noisy and scattered
about unity, theoretical considerations will give us a key to
understand and interpret the (small) deviations observed -- to be
discussed in the following sub-sections. We only briefly note here
that \ude\ haloes' parameter do not show, on average, any significant
sign of deviation from \LCDM.

\subsection{Virialization}
\label{sec:virialization}
It is known that the degree of virialization of dark matter haloes
with kinetic energy $K$ and potential energy $U$, which is usually
defined as
\begin{equation}
\left| \frac{U}{2 K} \right| = 1 \,
\end{equation}
is affected by the presence of an additional coupling \citep{Abdalla:2010, He:2010}.
In this case, due to the modification to the standard gravitational potential, 
the virial relation becomes
\begin{equation}\label{eq:vir_cde}
\left| \frac{U}{2 K} \right| = \frac{1 - \xi/2}{1 - 2\xi}
\end{equation}
where the parameter $\xi$ defined in \cite{Abdalla:2010}
can be written in terms of our definition of dark matter - dark energy coupling as:
\begin{equation}
\xi = \Omega_{dm} \beta_c.
\end{equation}

We can thus calculate the expected deviation from the standard
relation and compare it to the results of \Tab{tab:averages}. For
\cde099, this value is $1.04$, for \cde066 is $1.02$ while in \cde033
the value is $1.01$.
The predictions for these very small deviations from the \LCDM\ virial
equilibrium are compatible with the average findings of the
simulations presented in the upper panel of \Fig{img:cross_correlation},
although the large scatter
does not allow us to draw clear conclusions on the matter.  It
is however remarkable that, although weak, we can find evidence of
this modification.

\subsection{Spin parameter}
We use the spin parameter as defined by \citet{Bullock:2001}, i.e.
\begin{equation}
\lambda = \frac{L_{200}}{\sqrt{2}M_{200}V_{200}R_{200}}
\end{equation}
where the quantities $L$ (the total angular momentum), $M$ (total
mass), $V$ (circular velocity) and $R$ (radius) are computed using 
\Eq{eq:virial_mass_definition} with $\Delta=200$.
Our results (shown in the central panel of
\Fig{img:cross_correlation}) indicate that this parameter is
positively correlated to the coupling parameter $\beta_c$, as already
found in the analysis of smaller haloes in cosmologies where dark
matter feels an additional force (\cite{Hellwing:2011}, Paper I). 
For our models we find that $\lambda$ in \cde\
haloes differs on average up to a $\approx 10\%$ from its \LCDM\
cross-correlated partner, a result which is in good agreement 
with the findings of the aforementioned work.  

\subsection{Triaxiality}
We know that the shape of three dimensional haloes can be modelled
as an ellipsoidal distribution of particles \citep{Jing:2002, Allgood:2006}, 
whose three axes are given by the eigenvectors of the inertia tensor defined in \Eq{eq:itens_dm}.
Ordering the corresponding eigenvalues as $a\geq b\geq c$, we define the triaxiality parameter $t$ 
\footnote{We use $t$ instead of the commonly used $T$ to avoid any confusion with temperatures.}
as
\begin{equation}
t = \frac{a^2 - b^2}{a^2 - c^2}.
\end{equation}

In \Tab{tab:averages} we observe again a weak dependence of this
parameter on $\beta_c$ in \cde\ models.  \cde\ haloes here differ to
\LCDM\ correlated ones by values up to $6\%$. This effect is not
observed at lower masses (although not shown here), and -- like in the
previous case -- the scatter is quite large so that we definitely need
more statistics (i.e. simulations of larger volumes with the same mass
resolution) to ensure this is a real feature of massive dark matter
haloes in \cde\ models.

\subsection{$M_{200}$, $T_{mw}$ and $f_{gas}$}
The last halo properties we cross-correlated are mass,
gas fraction and mass weighted temperature, shown in the upper and lower panels of 
\Fig{img:cross_correlation}. 
The scatter in the first two is extremely small, with the ratios
clustering around unity; $T_{mw}$ on the other hand seem to vary more
across models even though still very close to one. Moreover, no sign
of dependence on the kind of quintessence or coupling seems to emerge.
So, even though we observed that gas and dark matter are distributed
differently, it is clear that the integral values of $M_{200}$ and
$f_{gas}$ cannot be used as a proxy for these discrepancies.  
It is interesting to note how the gas fraction, which we found to be strongly
correlated to the coupling parameter when projected radially, seems to 
be now unaffected by the interaction. However, this is not surprising, 
since a smaller abundance of gas in the central regions of the cluster is expected
to have a negligible effect on the overall $f_{gas}$, due to the little relative 
weight of the innermost regions. In a typical cluster, the gas mass at $r=0.1 \times R_{200}$
accounts for only a $3 - 4\%$ of the total, so that changes even as large
as $10\%$ only but slightly affect the global balance.
In any case, the histories of accretion of these parameters may indeed vary, even
bringing about the same results at $z=0$, as found by
\cite{Giocoli:2013} in the context of other coupled quintessence
models.  The behaviour of this quantities at higher redshifts will
be analyzed in an upcoming future work.

\section{Conclusions}\label{sec:conclusions}
In this contribution we have studied the properties of clusters and large
groups of galaxies using the set of hydrodynamical $N$-body
simulations introduced an earlier work (Paper I). The models under consideration
include a fiducial \LCDM\ cosmology, an uncoupled Dark Energy (\ude) 
and three coupled Dark Energy (\cde) models.
In each of them we have identified $\approx330$ structures with masses
$M_{200} > 7\times10^{13}$\hMsun\ which we further subdivided into
smaller subsets to best fit each one of our analysis purposes.  The
aim was to identify and possibly quantify systematic effects of
interacting (\cde) and non-interacting (\ude) quintessence on
properties of large and massive structures at $z=0$, and hence
providing a deeper understanding of the phenomenological consequences
arising in the non-linear regime of this class of theories. 

Our analysis was carried along two conceptually different lines,
namely investigating general properties of the set of objects, and then
one-to-one comparisons of cross-identified haloes. The first,
presented in \Sec{sec:clusters}, focused upon the determination of the
average properties of structures by considering homogeneous samples of
objects above a given mass cut. In this way we determined how
observables generally behave in different cosmologies.  While
integrated properties of the clusters, such as the X-ray
temperature-mass relation, tend to conceal any dependence on the
model, a closer look at the internal structure reveals that \cde\
models tend to favour less concentrated dark matter haloes and gas
fractions which are around $5\%$ below \LCDM\ in the innermost regions
of the clusters. We interpret this result as a consequence of the
reduction of dark matter density in the very same regions, which is
also proportional to the coupling. 
In our case, the suppression is $\approx10\%$, and is also related to an average
increase of the same magnitude of the peak value of the scale radii distribution.
This effect was described theoretically by \cite{Mainini:2005, Mainini:2006} and later found
in $N$-body simulations for galaxy groups and small clusters by \cite{Baldi:2010a, Li:2011}. 

The most remarkable findings, however, stem from the study of the
radial gas density and pressure profiles.  Although we have seen that
the extended $\beta$ model of \cite{Mroczkowski:2009} and the
observations of pressure profile of \cite{Arnaud:2010} seem to be able
to account for the numerical results to the same degree, \cde099 and
\cde066 still show large differences at the outer cluster edge. In
fact, since these properties are related to the ratio
$\rho_{gas}(r)/\rho_0$, due to the smaller $\rho_0$ the ratio becomes
larger when approaching $R_{200}$, and eventually leading to
discrepancies $>20\%$ for pressure profiles, which is so far the
largest difference predicted by us and for our models, respectively.

In addition, we have checked that the standard linear relation
for temperature profiles in the outskirts of the clusters holds also
in the case of \ude\ and \cde. Even the scatter in the determination
of the cluster mass under the hypothesis of hydrostatic equilibrium
seems to be largely model independent. However, it remains to be
confirmed whether these statements remain when taking into account a
larger sample of (even) more massive haloes.

Furthermore, in \Sec{sec:cross} we focused upon
individual structures and cross-correlated objects found in the \LCDM\
model to their counterparts in the other models.  This sort of
exercise is strictly theoretical and is aimed at determining which
properties of objects forming from comparable initial (Gaussian)
conditions and ending up at $z=0$ in similar environments are most
affected and thus likely to keep trace of the cosmological history.

We established that, whereas masses, total gas fractions and mass
weighted temperatures do not seem to be affected by cosmology,
virialization, spin parameter and triaxiality seem to be dependent on
the coupling in the dark sector, albeit only weakly. In particular, we
observed that deviations from the standard virial relations are in
agreement with the analytical values computed using the formula of
\cite{Abdalla:2010}, while spins seem to follow the pattern found in
Paper I at lower mass ranges. 

To conclude, we have examined the
impact of coupled dark energy in a series of galaxy group and cluster observables at
$z=0$. In some cases, we managed to establish a physical link between
the key observational properties and the underlying modified physical laws.  
However, it is still necessary to study the way background
quintessence and scalar field mediated interactions work at higher
redshifts and on larger and more massive structures. This will
be the focus of future contributions.

\section*{Acknowledgements}
EC is supported by the {\it Spanish Ministerio de Econom\'ia y
  Competitividad} (MINECO) under grant no. AYA2012-31101, and
MultiDark Consolider project under grant CSD2009-00064. 

AK is supported by the {\it Spanish Ministerio de Ciencia e
  Innovaci\'on} (MICINN) in Spain through the Ram\'{o}n y Cajal
programme as well as the grants CSD2009-00064,
CAM S2009/ESP-1496 (from the ASTROMADRID network) and the {\it
  Ministerio de Econom\'ia y Competitividad} (MINECO) through grant
AYA2012-31101. He further thanks Jasmine Minks for another age.

GY  acknowledges  support from  MINECO under research grants 
AYA2012-31101, FPA2012-34694,
Consolider Ingenio SyeC CSD2007-0050 and from Comunidad de Madrid under 
ASTROMADRID project (S2009/ESP-1496).

The authors  thankfully acknowledge the computer resources, technical expertise and assistance provided by the Red Espa\~nola de Supercomputaci\'on.

This work was undertaken as part of the Survey Simulation Pipeline (SSimPL: ssimpluniverse.tk)
and GFL acknowledges support from ARC/DP 130100117

We further acknowledge partial support from the European Union FP7 ITN
INVISIBLES (Marie Curie Actions, PITN-GA-2011-289442).

All the simulations used in this work were performed in the
Marenostrum supercomputer at Barcelona Supercomputing Center (BSC).


\bibliographystyle{mn2e}
\bibliography{biblio}

\bsp

\label{lastpage}

\end{document}